\documentstyle[12pt,axodraw]{article}
\begin{document}
\def\thebibliography#1{\section*{REFERENCES\markboth
 {REFERENCES}{REFERENCES}}\list
 {[\arabic{enumi}]}{\settowidth\labelwidth{[#1]}\leftmargin\labelwidth
 \advance\leftmargin\labelsep
 \usecounter{enumi}}
 \def\newblock{\hskip .11em plus .33em minus -.07em}
 \sloppy
 \sfcode`\.=1000\relax}
\let\endthebibliography=\endlist

\hoffset = -1truecm
\voffset = -2truecm

\title{\large\bf
Temperature Dependence of Gluon and Quark Condensates as from Linear
Confinement}
\author{
{\normalsize \bf
A.N. Mitra* \thanks{e.mail: ganmitra@nde.vsnl.net.in} \quad and
W-Y. P. Hwang \thanks{e.mail: wyhwang@phys.ntu.edu.tw}
}\\
\normalsize  Center for Academic Excellence on
Cosmology and Particle Astrophysics,\\
Department of Physics, National Taiwan University,\\
Taipei, Taiwan, R.O.C. \\
*Permanent address: 244 Tagore Park, Delhi-110009, India
}
\date{}
\maketitle


\begin{abstract}
The gluon and quark condensates and their temperature dependence are
investigated within QCD premises. The input for the former is a gauge
invariant $gg$ kernel made up of the direct (D), exchange (X) and contact(C)
QCD interactions in the lowest order, but with the perturbative propagator
$k^{-2}$ replaced by a `non-perturbative $k^{-4}$ form obtained via two differentiations:
$ \mu^2 \partial_m^2 (m^2+k^2)^{-1}$, ($\mu$ a scale parameter), and then setting $m=0$, 
to simulate linear confinement. Similarly for the input $q{\bar q}$ kernel the gluon 
propagator is replaced by the above $k^{-4}$ form. With these `linear' simulations, 
the respective condensates are obtained by `looping' up the gluon and quark lines in 
the standard manner. Using Dimensional regularization (DR), the necessary integrals yield 
the condensates plus temperature corrections, with a common scale parameter $\mu$ for 
both.  For gluons the exact result is
$$ <GG> = {36\mu^4}\pi^{-3}\alpha_s(\mu^2)[2-\gamma - 4\pi^2 T^2/(3\mu^2)]$$
Evaluation of the quark condensate is preceded by an approximate solution of the SDE for 
the mass function $m(p)$, giving a recursive formula, with convergence
achieved at the third iteration. Setting the scale parameter $\mu$ equal to
the universal Regge slope $1 GeV^2$, the gluon and quark condensates at $T=0$
are found to be $0.586 Gev^4$ and $(240-260 MeV)^3$ respectively, in fair
accord with QCD sum rule values. Next, the temperature  corrections (of order $-T^2$ for both 
condensates) is determined via finite-temperature field theory a la Matsubara.  \\
Keywords: Gluon Condensate, mass tensor, gauge invariance, linear
confinement, finite-temperature, contour-closing. \\
PACS: 11.15.Tk ;  12.38.Lg ; 13.20.Cz  
\end{abstract}

\section{Introduction}

The thermal behaviour of QCD parameters has acquired considerable
relevance in recent times in the context of global experimentation
on heavy ion collisions as a means of accessing the quark-gluon
plasma (QGP) phase. There are two distinct aspects to the effects
of finite temperature, viz., deconfinement and restoration of chiral
symmetry at high enough temperatures, but the question of which one
of these two should occur before the other, is probably one of
technical details of the theory [1]. We are concerned here
not so much with the similarity of these phenomena as with the general
manner of their onset, to the extent that the temperature variation of
certain condensates carries this information. Indeed it is generally believed
that the thermal behavour of condensates provides a fairly reliable index to
the deconfinement/chiral symmetry restoration phases, but the precise
mechanism through which this occurs often gets obscured in the
technical details of the models employed.
To that end, the quark and gluon condensates, by virtue of their basic nature,
offer themselves as prime candidates  for a study of the problem, to
the extent that such phase transitions  can be inferred from their
thermal behaviour. In this respect, QCD-SR has been a leading candidate for such
investigations for more than two decades [2], yet it cannot be regarded as a full-fledged
substitute for a non-perturbative treatment in a $closed$ form since the OPE underlying 
it stems from the high energy end. Prima facie it appears that the QCD-SR- prediction 
of temperature correction to $<GG>$  of order $T^4$ [3,4], instead of $T^2$, is rather 
`flat' unless justified by more convincing arguments. On the other hand, the chiral 
perturbation theory, with non-perturbative information of a different type built-in [5], 
gives temperature corrections of $O(T^2)$ for $<q{\bar q}>_0$ [6]), while  the non-perturbative 
content of instanton theories [7] is of a different nature. In view of such wide variations in 
the different predictions it should be worth exploring still other non-perturbative 
approaches to QCD, especially ones which incorporate $confinement$ more transparently.

\setcounter{equation}{0}
\renewcommand{\theequation}{1.\arabic{equation}}

        With this idea in mind, we propose an alternative non-perturbative approach
to QCD and test it initially against the problem of the two basic QCD condensates. The
precise content of this approach is expressed by the replacement of the o.g.e.
propagator $k^{-2}$ by a non-perturbative form $k^{-4}$ which corresponds to 
$linear$ confinement (in an `asymptotic' sense) according to conventional wisdom, as follows:
\begin{equation}\label{1.1}
lim_{m \rightarrow 0}{(\mu \partial_m)}^2 \frac{1}{k^2+m^2} = \frac{2\mu^2}{k^4}
\end{equation}
where $\mu$ is a scale parameter to be specified further below. While 
the full ramifications of this form are still being developed (see Section 6), the present study is
limited only to its predictions on the condensates $<GG>$ and $<q{\bar q}>$ plus $f_\pi$.
Consider first the gluon case where the dominant interaction is provided by the gluons 
themselves, with quark effects playing a secondary role. Now the standard QCD Lagrangian 
defines both the 3-and 4-gluon vertices, from which the lowest order $gg$ interaction, 
via gluon exchange (both the `direct'(D) and `exchange' (X) terms), as well as
the 4-gluon `contact'(C) term can be generated as a {\it gauge invariant}
package proportional to the QCD coupling constant $\alpha_s(\mu)$ [8,9](see fig.1):
\begin{equation}\label{1.2}
\frac{g_s^2 (\mu)F_1.F_2}{k^2}(1+P_{12})[V^{(1)}.V^{(2)}-\frac{k^2}{2}
\Delta(12)]
\end{equation}
where $P_{12}$ a permutation operator, and the
other symbols are defined later in Sect. 2.  This structure was employed in [8,9]
as a kernel of a BSE for the $gg$ wave function for the calculation of
glueball spectra, but with the perturbative propagator replaced by harmonic
type confinement, on similar lines to $q{\bar q}$ spectroscopy [10]. 
Alternative BSE treatments for glueballs also exist in the literature [11]. In this
paper, on the other hand, we seek specifically to examine the precise point of
departure of Eq.(1.2) from perturbative QCD by replacing the o.g.e propagator $k^{-2}$ in 
front by its $non-perturbative$ form (1,1), (with a conscious simulation of $linear$ confinement), 
while retaining both Lorentz and gauge invariance of the full ($D+X+C$) package. This $gg$
interaction is proposed as a candidate for non-perturbative treatment of a few leading
QCD amplitudes, with the burden of variation from perturbative QCD resting
on the mathematical properties of a simple differential operator. Among possible 
applications,this interaction  may serve as the kernel of a BSE on the lines of (8-11), but even in
the $lowest$ order this non-perturbative quantity also offers other interesting possibilities.
For, in the absence of quark (fermion) interactions, there is no coupling of the (vector) gluon 
field to $axial$ currents [12], hence no possibility of  `poles'  in the 
gluon self-energy tensor [12] via the Schwinger mechanism [13]. Therefore it
makes sense to calculate the gluon self-energy tensor by `looping up' two of the gluon
lines, and as a next step, the gluon condensate by letting the other two gluon
lines disappear into the vacuum, a la Fig. 2. The derivative
structure of Eq. (1.1-2) is ideally suited to the 't Hooft- Veltman method [14] of
dimensional regularization (DR), while the temperature dependence a la Matsubara [15] may be
introduced by an elegant method due to Kislinger-Morley [16].
The final result for the gluon condensate, including temperature effects, is
extremely simple:
\begin{equation}\label{1.3}
 <GG> = \frac{36 \mu^4}{\pi^3}(2-\gamma)\alpha_s(\mu^2)[1- (2\pi T/\mu)^2/(6-3\gamma)]
\end{equation}
where $\gamma$ = $0.5772$ is the Euler-Mascheroni constant. It is
rather amusing that this value $(0.586 GeV^4)$ comes fairly close to
the QCD-SR value of (0.480) [3] if the DR parameter $\mu^2$ (the only free parameter
in the theory) is identified with the universal Regge slope of $~1 GeV^2$. However the
temperature correction of $O(T^2)$ seems to be larger than the QCD-SR value [4], while
comparison with other models is postponed till Section 6. 

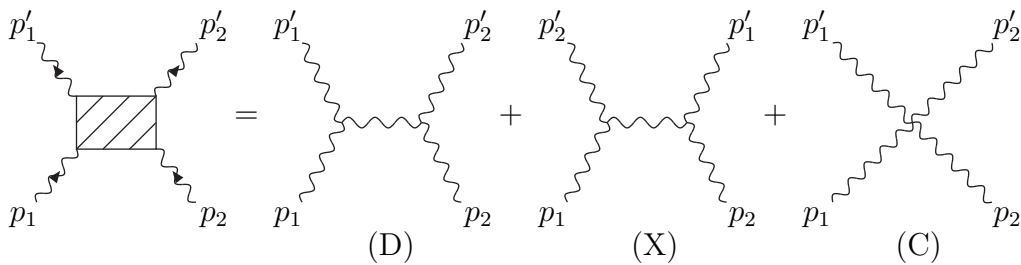
\begin{figure}
\begin{picture}(400,110)(0,0)
\BBox(40,40)(70,60)
\Photon(25,20)(40,40){2}{3}
\Photon(25,80)(40,60){2}{3}
\Photon(85,20)(70,40){2}{3}
\Photon(85,80)(70,60){2}{3}
\ArrowLine(32.5,30)(32.515,30.020)
\ArrowLine(32.515,70)(32.5,70.020)
\ArrowLine(77.515,30)(77.5,30.020)
\ArrowLine(77.5,70)(77.515,70.020)
\Line(50,40)(70,60)
\Line(40,40)(60,60)
\Line(60,40)(70,50)
\Line(40,50)(50,60)
\put(15,13){$p_1$}
\put(15,84){$p_1^{\prime}$}
\put(87,13){$p_2$}
\put(87,84){$p_2^{\prime}$}
\put(100,50){$=$}
\Photon(125,20)(140,50){2}{4}
\Photon(125,80)(140,50){2}{4}
\Photon(185,20)(170,50){2}{4}
\Photon(185,80)(170,50){2}{4}
\Photon(140,50)(170,50){2}{3}
\put(115,13){$p_1$}
\put(115,84){$p_1^{\prime}$}
\put(187,13){$p_2$}
\put(187,84){$p_2^{\prime}$}
\put(200,50){$+$}
\Photon(225,20)(240,50){2}{4}
\Photon(225,80)(240,50){2}{4}
\Photon(285,20)(270,50){2}{4}
\Photon(285,80)(270,50){2}{4}
\Photon(240,50)(270,50){2}{3}
\put(215,13){$p_1$}
\put(215,84){$p_2^{\prime}$}
\put(287,13){$p_2$}
\put(287,84){$p_1^{\prime}$}
\put(300,50){$+$}
\Photon(325,20)(385,80){2}{10}
\Photon(325,80)(385,20){2}{10}
\put(315,13){$p_1$}
\put(315,84){$p_1^{\prime}$}
\put(387,13){$p_2$}
\put(387,84){$p_2^{\prime}$}
\put(150,0){(D)}
\put(250,0){(X)}
\put(350,0){(C)}
\end{picture}
\caption {
A gauge invariant package of $gg$ interaction in second order of QCD.
D, X, C represent direct, exchange and contact interactions respectively.}
\end{figure}$\;$
\begin{center}
\begin{figure}
\begin{picture}(310,110)(0,0)
\BBox(50,40)(110,60)
\Photon(35,20)(50,40){2}{4}
\Photon(135,20)(110,40){2}{4}
\ArrowLine(42.5,30)(42.515,30.020)
\ArrowLine(122.515,30)(122.5,30.020)
\PhotonArc(80,60)(30,0,180){2}{10}
\Line(50,40)(70,60)
\Line(60,40)(80,60)
\Line(70,40)(90,60)
\Line(80,40)(100,60)
\Line(90,40)(110,60)
\Line(100,40)(110,50)
\Line(50,50)(60,60)
\put(22,13){$p$}
\put(138,13){$p$}
\put(65,95){$p-k$}
\BBox(250,40)(310,60)
\PhotonArc(280,60)(30,0,180){2}{10}
\PhotonArc(280,40)(30,180,360){2}{10}
\ArrowLine(280.030,10)(280,10)
\Line(250,40)(270,60)
\Line(260,40)(280,60)
\Line(270,40)(290,60)
\Line(280,40)(300,60)
\Line(290,40)(310,60)
\Line(300,40)(310,50)
\Line(250,50)(260,60)
\put(265,95){$p-k$}
\put(310,13){$p$}
\put(0,50){(a)}
\put(200,50){(b)}
\end{picture}
\caption {
Looping up of gluon lines to give
(a) self-energy tensor and (b) gluon condensate.}
\end{figure}
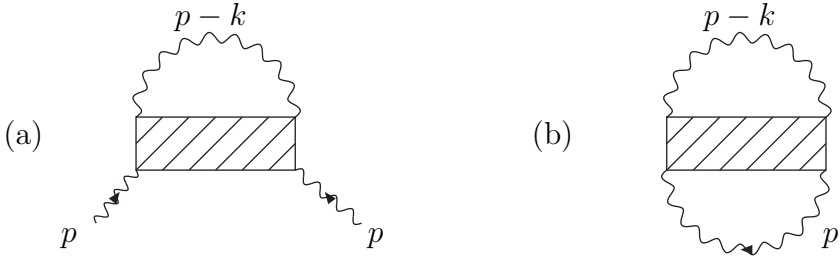
\end{center}

As  a second test of our non-perturbative ansatz (1.1), we shall consider the $q{\bar
q}$ interaction via o.g.e., wherein the factor $g_s^2 F_1.F_2/k^2$ in Eq. (1.2) is 
replaced by $g_s^2 f_1.f_2 \mu^2/k^4$, and generate the mass function $m(p)$ via SDE, 
and thence  the quark condensate [17], via this new non-perturbative interaction. A
third test is to the pionic constant $f_\pi$ by the same method.    
In Sect. 2 we give a derivation of the analytic structure of the
gluon condensate based on the diagrams of Figs. 1 and 2. In Sect. 3 we
evaluate this quantity a la ref. [14], with explicit gauge invariance
built in, and also introduce the temperature dependence a la ref. [16],
to obtain Eq. (3). Sects. 4 and 5 give a corresponding derivations of the quark
condensate and pionic constant respectively, showing equally promising results with the 
same scale parameter $\mu=1GeV$. Sect. 6 concludes with a short discussion on the
non-perturbative ansatz (1.1) vis-a-vis the more standard approaches.  

\section{Structure of the Gluon Condensate}

\setcounter{equation}{0}
\renewcommand{\theequation}{2.\arabic{equation}}

We first spell out the structure of the $gg$ scattering amplitude
arising from the interaction (1.1-2), in accordance with the figures
1(a), 1(b), 1(c) corresponding to direct (D), exchange (X) and
contact (C) processes respectively. To that end the momenta and polarization
indices of the incoming gluon lines are  denoted by $(p_1,\mu_1; p_2,\mu_2)$,
and of the outgoing ones by $p_1',\nu_1; p_2',\nu_2$ respectively [8].
In this notation, the various quantities are expressed as follows [8]:
\begin{equation}\label{2.1}
V_\mu^{(1,2)}  = (p_{1,2}+p'_{1,2})_\mu +2i S_{\mu\nu}^{(1,2)}(p'_{1,2}-p_{1,2})_\nu; \quad
-i [S^{(1)}_{\mu\nu}]_{\mu_1\nu_1} = \delta_{\mu\mu_1}\delta_{\nu\nu_1} - \delta_{\mu\nu_1}\delta_{\mu_1\nu}
\end{equation}
and a similar value for $S^{(2)}_{\mu\nu}$. The spin dependence of the contact term is given by [8]:
\begin{equation}\label{2.2}
\Delta^{\nu_1\nu_2}_{\mu_1\mu_2} = \delta_{\mu_1\nu_1} \delta_{\mu_2\nu_2} + \delta_{\mu_1\nu_2} \delta_{\mu_2\nu_1}
 - 2 \delta_{\mu_1\mu_2} \delta_{\nu_1\nu_2}
\end{equation}
The permutation operator $P_{12}$ in Eq. (1.2) provides the exchange process by
simultaneously interchanging these collective indices for the $outgoing$ lines,
while holding the incoming one fixed. However, unlike the case of glueball
spectra [8-9], the effect of the 
permutation operator is not merely a factor of two, but a non-trivial sum of
two different groups of terms related by the above interchanges. Defining the
total ($P$) and relative ($q,q'$) momenta as [8,9]
\begin{equation}\label{2.3}
P=p_1+p_2 =p_1'+p_2'; \quad 2q =p_1-p_2; \quad 2q'=p_1'-p_2'; \quad P^2 = -M^2
\end{equation}
The momentum and spin dependence of the `direct' part plus $half$ the `contact'
part of (1.2)may now be adapted from [8] to give
\begin{eqnarray}\label{2.4}
\lefteqn{A(p_1,p_2; p_1'p_2')_{\mu_1\mu_2}^{\nu_1\nu_2} =}  \\ \nonumber
& & [-M^2-(q+q')^2] \delta_{\mu_1\nu_1}\delta_{\mu_2\nu_2} +[(p_1+p_1')_{\mu_2}(-2p_2)_{\nu_2}+ \\ \nonumber
& & (p_1+p_1')_{\nu_2}(-2p_2')_{\mu_2}] \delta_{\mu_1\nu_1} +[(p_2+p_2')_{\mu_1}(-2p_1)_{\nu_1}  \\ \nonumber
& & +(p_2+p_2')_{\nu_1}(-2p_1')_{\mu_1}] \delta_{\mu_2\nu_2} +(-2p_1)_{\nu_1}(-2p_2)_{\nu_2}\delta_{\mu_1\mu_2}   \\  \nonumber
& & +(-2p_1)_{\nu_1}(-2p_2')_{\mu_2}\delta_{\mu_1\nu_2}  +(-2p_1')_{\mu_1}(-2p_2)_{\nu_2}\delta_{\nu_1\mu_2}  \\  \nonumber
& &  +(-2p_1')_{\mu_1}(-2p_2')_{\mu_2}\delta_{\nu_1\nu_2}- \frac{1}{2} (q-q')^2 \Delta^{\nu_1\nu_2}_{\mu_1\mu_2}
\end{eqnarray}
The corresponding `exchange' part is then given by $ A(p_1,p_2;
p_2'p_1')_{\mu_1\mu_2}^{\nu_2\nu_1}$. We now tie up the two legs $p'_1,\nu_1$
and $p_2',\nu_2$ in the above expression into a propagator (in the simple 
Feynman gauge), in accordance with Fig. 2(a), together with a similar
expression for the corresponding exchange part. Before writing this down, we relabel the momenta and polarizations in (2.4) as follows
$$ p_1 =-p_2 = p; \quad p_1' =-p_2' = p-k; \quad q+q'=2p-k; \quad q-q'= k;
\quad \mu_1,\mu_2= \mu,\nu$$ 
For the exchange part, simply change the sign of $q'$ which will interchange
$k$ with $2p-k$, and 
$\nu_1$ with $\nu_2$. The resultant quantity  simplifies after several vital
cancellations to 
\begin{equation}\label{2.5}
N_{\mu\nu}(p,k) = [(2M^2 + 2k^2 + 2(2p-k)^2)\delta_{\mu\nu}- 8(p-k)_\mu (p-k)_\nu
 + 8(p_\mu p_\nu- p^2 \delta_{\mu\nu})
\end{equation}
which, apart from constant factors, can be related to the gluon self-energy tensor $\Pi_{\mu\nu}(p)$ by
\begin{eqnarray}\label{2.6}
\Pi_{\mu\nu}(p) &=& [\mu^2 \partial_m^2 ]_{m=0}]
\int \frac{-i d^4k}{(2\pi)^4} \frac{N_{\mu\nu}(p,k)}{(m^2+ (p-k)^2)(m^2+k^2)} \\  \nonumber
                & &  \equiv (p_\mu p_\nu-p^2\delta_{\mu\nu}) \Pi(p^2)
\end{eqnarray}
We have used the same mass for the propagator of the internal gluon line $p-k$.
Since this quantity must be explicitly gauge invariant, we have anticipated its
proportionality to $(p_\mu p_\nu-p^2\delta_{\mu\nu})$ in the second line. The
final stage involves the looping up of the remaining two gluon lines into the
vacuum, in the standard fashion [18] which includes taking a 4D `curl' of the
(non-perturbative) gluon propagator to get $<G_{\mu\nu} G_{\nu\mu}>$. These
routine operations on Eq. (2.6) yield 
\begin{equation}\label{2.7}
<G^a_{\mu\nu} G^b_{\nu\mu}> = [6 C] \delta^{ab} \int \frac{-i d^4p}{(2\pi)^4} \Pi (p^2)]
\end{equation}
where $\Pi(p^2)$ is defined in Eq. (2.6). The
factor $[6]$ in front arises from the effect of `curl' on the gluon propagator
to obtain the LHS, while $[C]$ represents the remaining constants from the
lowest order QCD amplitude of Eq. (1.1-2), viz., 
\begin{equation}\label{2.8}
 C= 4\pi \alpha_s(\mu^2) F_1.F_2
\end{equation}
$F_1.F_2 =-3$ being the color Casimir of the $g-g$ interaction [9] which should be distinguished from
the factor $\delta^{ab}=8$ arising from the last gluon propagator before disappearing into the vacuum.

\section{Dimensional Regularization for Integrals}

\setcounter{equation}{0}
\renewcommand{\theequation}{3.\arabic{equation}}

We first consider the integral of Eq. (2.6) for which we shall follow closely the
DR method [14]. Introducing 
the Feynman variable $0 \leq u \leq 1$, giving the usual shift $k\rightarrow p
+k u$ and dropping the odd terms in $k_\mu$ upstairs, the resultant integral
reduces to 
\begin{eqnarray}\label{3.1}
\Pi_{\mu\nu}  &=& [\mu^2 \partial_m^2 ]_{m=0}]\mu^{(4-n)}\int \frac{d^n k}{(2\pi)^n} \int_0^1 du  \nonumber \\
              & & \frac{[(8p_mu p_\nu -8 p^2\delta_{\mu\nu})(2u-u^2)
 +(4 p^2 (2u-u^2)-2M^2 +2k^2)\delta_{\mu\nu}]}{[m^2 + p^2 u(1-u)+k^2]^2}
\end{eqnarray}
The result of DR may be expressed by the formula [14]:
\begin{equation}\label{3.2}
\mu^{(4-n)} \int d^n k \frac{[k^2; 1]}{[k^2 + A]^2} = \frac{\pi^{n/2}\Gamma(2-n/2)}{(A/\mu^2)^{(2-n/2)}}[\frac{n}{2-n}; 1]
\end{equation}
Subtracting in the usual manner the residue-cum-pole parts corresponding to
$n=4$, it is easily seen that gauge invariance is satisfied except for the
portion  $2(p^2 - M^2)\delta_{\mu\nu}$ in Eq. (3.1), which vanishes for
$p^2=M^2$ only! We are unable to offer an interpretation of it, but we see
little alternative to `regularizing' it before proceeding further. The final result
after taking out $(p_\mu p_\nu-p^2\delta_{\mu\nu})$ is 
\begin{equation}\label{3.3}
\Pi (p^2) = [\mu^2 \partial_m^2]\int_0^1 du 8(1/2 +
u(1-u))\frac{1}{2\pi^2}[\gamma + ln (A/\mu^2)]; \quad  A= m^2 +p^2 u(1-u)
\end{equation} 
where we followed the precaution advocated in ref. [14], viz., dropping terms
odd in $(2u-1)$, leading to the replacement $u(2-u)\rightarrow$ $1/2 + u(1-u)$
in the numerator of Eq. (3.3). 

\subsection{ Gluon Condensate at $T=0$}

The gluon condensate may be obtained by substituting Eq. (3.3) in Eq. (2.7), where
the second derivative w.r.t. $m$ (multipled by $\mu^2$), gives
\begin{equation}\label{3.4}
<G^a G^b> = [6 C] \delta^{ab} \int \frac{-i d^4p}{(2\pi)^4}\int_0^1 du \frac{(1/2 + u(1-u))}{\pi^2[m^2+ p^2 u(1-u)]}
\end{equation}
Once again the $p$-integration is done a la DR [14], Eq. (3.2) and the result of
$p$-integration is 
\begin{equation}\label{3.5}
<G^a G^b> = [6 C] \delta^{ab} \mu^2 \int_0^1 \frac{du m^2}{u^2(1-u)^2 \pi^2}(1/2 +u(1-u))[\gamma-1 - ln(u \mu^2/ m^2)]
\end{equation}
The $u$-integration is now routine, and in the limit $m=0$ only the $(1/2)$
part survives, and as a check on the consistency of the calculation, the
$ln(\mu^2/m^2)$ cancel out exactly. The final result using Eq. (2.8) is 
\begin{equation}\label{3.6}
<GG>_0   = \frac{[36\mu^4\alpha_s(\mu^2)]}{\pi^3}[2-\gamma]
\end{equation}
where the common scale parameter  $\mu^2$ has been employed throughout.  If one
uses the universal Regge slope value of $\mu^2 \approx 1GeV^2$, then $<GG>_0=
0.586GeV^4$, slightly on the high side of the QCD-SR value [3], without any
fine-tuning. 

\subsection{ Temperature Dependence of Gluon Condensate}

So far our formalism  has been fully Lorentz and gauge invariant. To
incorporate the temperature dependence of $<GG>$, a convenient starting point
is the quantity $\Pi(p^2)$, Eq. (3.3), in which the time-like component $p_0$
is analytically continued, a la Matsubara [15], to the imaginary axis, with
discrete eigenvalues $ i p_0 = 2\pi T n$ (for bosons). The $p$-dependence of
$\Pi(p^2)$ may be recognized through the denominator $(m^2 + u p^2)^{-1}$ in
the integral for $<GG>$ in Eq. (3.4) where the substitution $p^2 \rightarrow
{\hat p}^2 + (2\pi T n)^2$ will give  a `discreteness' to the time-like
momentum. To evaluate $<GG>$ via Eq. (3.4), we now convert it into a contour
integral over $p_0$, a la Kislinger-Morley [16]. To that end we reproduce their
Eq. (2), except for $K^0 \rightarrow p_0$: 
\begin{eqnarray}\label{3.7}
2i\pi T \sum f(\nu_n) &=& \int_{-i\infty+\epsilon}^{+i\infty+\epsilon} \frac{dp_0 f(p_0)}{\exp{(p_0/T)}-1} \\ \nonumber
                      & & + \int_{-i\infty-\epsilon}^{+i\infty-\epsilon} \frac{dp_0 f(p_0)}{\exp{(-p_0/T)}-1} +\int dp_0 f(p_0)
\end{eqnarray}
where the last term (independent of $T$) is precisely identifiable with our
Eq. (3.4) $after$ integration 
over $d^3{\hat p}$. On the other hand,  the first two terms which are
temperature dependent, are most easily evaluated by their method of `contour
closing' [16] in the variable $p_0$, which amounts to the replacement of $p_0$
by its `pole' values $\pm \omega_u$ = $\pm \sqrt{ (m^2/u(1-u) +{\hat p}^2)}$ in
the two Bose-Einstein functions respectively, giving exactly equal
contributions. [This result agrees with the one found in ref. [16] for the
one-loop temperature-dependent correction to $mass^2$ in the $\phi^4$ theory,
since the analytic structures of both are similar]. As to the $u$-integration,
the pieces $1/2$ and $u(1-u)$ have rather complementary roles: The former had
given a finite non-zero limit Eq. (3.6) for $T=0$, only due to DR [14], but now
in the absence of (a second stage) DR for $T>0$, this term gives a divergence
for $m=0$, and there is little alternative to its "regularization". On the
other hand the $u(1-u)$ term which had vanished for the $T=0$ case, now gives a
perfectly finite value for $T>0$ in the $m=0$ limit. The remaining integrations
are all convergent and elementary on setting $m=0$ at the outset, so that the 
$T$-dependent part of $<GG>$, via Eq. (3.4) becomes:
\begin{equation}\label{3.8}
<GG>_T = [4C] \frac{\mu^2 T^2}{\pi^2} = -<GG>_0 \frac{4\pi^2 T^2}{\mu^2 (6-3\gamma)}
\end{equation}

\section{ The Quark Mass Function and Condensate}

\setcounter{equation}{0}
\renewcommand{\theequation}{4.\arabic{equation}}

We now come to our next item, the quark condensate, which however requires a
prior determination of the mass function as the crucial ingredient. Thus must
be done dynamically 
via the(non-perturbative) SDE for which we follow the treatment of ref. [17],
now adapted to the confining interaction of Eq. (1.1):
\begin{equation}\label{4.1}
m(p) = 3g_s^2 f_1.f_2 [\mu^2 \partial_m^2] \int \frac{-i d^4 k}{(2\pi)^4} \frac{m(p-k)}{(m^2+k^2)[m^2(p-k)+(p-k)^2]}
\end{equation}
where $g_s^2 = 4\pi \alpha_s$, $f_1.f_2 = -4/3$ is the color Casimir; the
Landau gauge has been employed [19], and $m=0$ after differentiation. Our 
defence of the Landau gauge is essentially one of practical expediency, since this  
gauge usually offers the safest and quickest route to a gauge invariant result, 
even without a detailed gauge check, for there has been no conscious violation 
of this requirement at any stage in the input assumptions. For the solution of   
this equation, unfortunately no exact analytic solution is available in this
case (unlike the previous case of harmonic confinement [17]), so we have developed 
an iterative analytical procedure as follows. As a first step, we replace the mass 
function inside the integral by $m(p)$.  Then the DR method [14] of Sect.3 may be 
used almost verbatim to evaluate the integral on the RHS of Eq. (4.1), by subtracting the
pole contribution and carrying out the indicated differentiation w.r.t. $m$.
Eq. (4.1) now reads : 
$$ m(p) = -3 f_1.f_2 m(p) \frac{\pi^2 g_s^2}{(2\pi)^4} \frac{\mu^2}{m^2(p)/2 + p^2/4} $$
which gives a first iterative solution (valid for momenta $p^2 \leq \mu^2$):
\begin{equation}\label{4.2}
m^2(p) = m_0^2 - \frac{1}{2} p^2; \quad m_0^2 \equiv m_q^2 = \frac{\alpha_s \mu^2}{\pi}
\end{equation}
after substituting for $f_1.f_2 = -4/3$ and using $g_s^2 = 4\pi \alpha_s$.
Then the first iteration for the self-energy operator is
\begin{equation}\label{4.3}
\Sigma_1(p) = \frac{-ig_s^2 f_.f_2}{(2\pi)^4} \int d^4 k \frac{\gamma_\mu (m_1-i\gamma.(p-k))\gamma_\mu}
{(m^2+k^2)[(m_1^2+(p-k)^2)/2]}
\end{equation}
where the second factor in the denominator on the RHS is the result of
substitution of the first iteration Eq. (4.2) for the mass function in
$m^2(p-k)+(p-k)^2$, giving $(m_1^2+(p-k)^2)/2$, 
and $m_1^2 = 2m_0^2$ represents the first iteration to $m_0^2$, Eq. (4.2), which
also goes into the numerator of Eq. (4.3). Now proceed exactly as in Sect. 3:
introduce the Feynman variable $u$, symmetrize wrt $u \leftrightarrow 1-u$,
integrate via DR [14], and differentiate twice wrt $m$. Then up to $O(p^2)$,
the $u$ integration gives: 
\begin{equation}\label{4.4}
\Sigma_1(p) = \frac{2m_0^2}{3} \frac{4 m_1+ i \gamma.p}{(m_2^2 + p^2)/4}; \quad m_2^2 = 2m_1^2=4m_0^2
\end{equation}
To account for the factor $1/4$ in the denominator on right, a factor $2$ each
is absorbed in the two flanking $S_{F1}(p)$ functions which are needed to
define the  quark condensate, in this order of iteration, which is given by
\begin{equation}\label{4.5}
<q{\bar q}>_1 = N_c Tr \int d^4 p S_{F1}(p) \Sigma_1(p) S_{F1}(p); \quad S_{F1}=(m_1+i\gamma.p)^{-1}
\end{equation}
where $\Sigma_1(p)$ of Eq. (4.4) must now be read without the said factor $4$.
This structure is quite general and reproduces itself at every stage of
iteration, which however must stop when the iterated mass $m_n$ gets near
$\mu$, due to the approximate nature of Eq. (4.2). Eq. (4.5) now simplifies to:
\begin{equation}\label{4.6}
<q{\bar q}>_1 = \frac{m_q^2}{4\pi^2} \int \frac{-i d^4 p 2m_1}{(2\pi)^4}\frac{m_2^2-p^2}{(m_1^2+p^2)^2(m_2^2+p^2)}
\end{equation}
The recursive law is now clear: In the next step, $<q{\bar q}>_2$ will be given
by replacing $m_1, m_2$ 
in Eq. (4.6) with $m_2,m_3$ respectively, where $m_3^2= 2m_2^2$; see Eq. (4.4).
And so on till 
the `limit' $\mu$ is reached. Evaluation of (4.6) by the DR method [14] now
yields the result: 
\begin{equation}\label{4.7}
<q{\bar q}>_1 = \frac{m_q^2 m_1}{2\pi^2} \int [4/3 + \gamma - ln(\mu^2/m_1^2)]
\end{equation}
from which the successive iterations can be directly written down (including
the `zero' order) since the `mass' increases by $\sqrt{2}$ at each step.
The numerical values (using the universal $\mu=1 GeV$ as before) for the 0,1,2
iterations are $(132 MeV)^3$, $(204 MeV)^3$ and $(270 MeV)^3$ respectively,
beyond which our simple formula (4.2) breaks down. 

\subsection{ Temperature Corrections to Quark Condensate}

To calculate the temperature dependence of the quark condensate, a convenient
starting point is Eq. (4.6), where the formalism of ref [16] applies a la Eq.
(4.7), with the function $f(p_0)$ identified as
$$ \frac{m_2^2-p^2}{(m_1^2+p^2)^2(m_2^2+p^2)};  \quad p^2 = {\hat p}^2 - p_0^2 $$
As in the gluon case the last term of (3.7) will give the $T=0$ condensate
(4.6), while the contour closing in the temperature dependent integrals will
give equal contributions, with the Bose-Einstein functions replaced by the
Fermi-Dirac functions. The contour closing now gives rise to the residues at
the poles $ p_0 = \omega_1$ (double) and $p_0 = \omega_2$ (single)
respectively, where $\omega_n^2 = m_n^2 + {\hat p}^2 $. Substituting the
residues and simplifying yields the net 3D integral: 
\begin{equation}\label{4.8}
<q{\bar q}>_T = \frac{10 m_q^2}{4\pi^2 m_2}\int \frac{d^3 p}{(2\pi)^3}
[g(\omega_2/T)-g(\omega_1/T)] 
\end{equation}
where
\begin{equation}\label{4.9}
g(y) = \frac{1}{y} \frac{1}{e^y +1}
\end{equation}
To evaluate (4.8), the substitutions ${\hat p}=Tx$ and $\omega_n = T \sqrt{x^2+\lambda_n^2}$,
where $m_n=T\lambda_n$, lead to the result:
\begin{equation}\label{4.10}
<q{\bar q}>_T = \frac{10 m_q^2 T^2}{\pi m_2}[F(\lambda_2)-F(\lambda_1)]
\end{equation}
where
\begin{equation}\label{4.11}
F(\lambda)= \int_0^\infty \frac{x^2 dx}{\sqrt{\lambda^2 + x^2}}[1+\exp{\sqrt{\lambda^2+x^2}}]^{-1}
\end{equation}
This integral can be rapidly evaluated by the substitution $x=\lambda cosh(\theta/\lambda)$,
and noting that terms of $O(\theta^2)$ (gaussian) give the main contributions. The
final result for $F(\lambda)$ is
\begin{equation}\label{4.12}
F(\lambda) \approx \sqrt{\pi \lambda/2} e^{-\lambda} [1- \frac{\exp{(-\lambda)}}{\sqrt{8}}]
\end{equation}
Substitution of numerical values shows that unlike the gluon case, the decrease with temperature
is rather slow,  being less than half percent at $150 MeV$, despite the $T^2$ dependence of
the correction, as in chiral perturbation theory [6]. 

\section{ Calculation of the Pionic Constant }

\setcounter{equation}{0}
\renewcommand{\theequation}{5.\arabic{equation}}

As a third and final test of this formalism, the evaluation of the pionic constant requires only an extra
ingredient which stems from the Ward identity for chiral symmetry breaking [12], viz., in the
`chiral limit' (i.e., when the pion mass vanishes), the self-energy operator $\Sigma(p)$
and the pion-quark vertex function $\Gamma{p_1,p_2}$ are identical [20-22]. Although originally
discovered for the contact interaction [20], its validity  was found for more general interactions
[21] which allowed an alternative derivation of the mass function by the Bethe-Salpeter route for
the pion-quark vertex function [17]. In a more elegant fashion, the content of this identity
was expressed by Pagels-Stokar [22] as follows:
\begin{equation}\label{5.1}
2 f_\pi \Gamma(p_1,p_2) = \Sigma(p_1) + \Sigma(p_2); \quad f_\pi = 93 (MeV)
\end{equation}
where for each $p_i$, $\Sigma(p)$ is given by (4.3) and simplifies to (4.4) after $k$-integration.
The pionic constant is formally defined (with $P=p_1+p_2$) [17] by
\begin{equation}\label{5.2}
f_\pi P_\mu = \int \frac{-i N_c d^4 q}{(2\pi)^4} Tr [\gamma_5 \Gamma(p_1,p_2)S_F(p_1)i\gamma_\mu \gamma_5 S_F(-p_2)]
\end{equation}
where $p_{1,2}=P/2 \pm q$, and $S_F(p_i)$ are given by eq.(4.5) in the same order of iteration as in Sect.4.
The trace evaluation is routine, allowing for calcellation of $P_\mu$ from both sides. Further, in the $P^2=0$ limit,
the integrand simplifies greatly to give (with $N_c=3$):
\begin{equation}\label{5.3}
f_\pi^2 = 8m_q^2 \int \frac{-i d^4 q}{(2\pi)^4} \frac{3m_1^2 + (m_1^2 +q^2)}{[(m_2^2 +q^2)(m_1^2 +q^2)^2]}
\end{equation}
where the various symbols are the same as in Sect 4. The first term on the RHS is convergent as
it is, while the second one amenable to a DR treatment [14], exactly as in the two previous
sections.. The final result, in the chiral limit $P^2 = 0$ is
\begin{equation}\label{5.4}
f_\pi^2 = \frac{m_q^2}{2\pi^2} [1 - \gamma + ln(\mu^2/3m_q^2)]
\end{equation}
where $m_q$ is related to $m_1$ and $m_0$ by $m_1^2=2m_0^2=2m_q^2$, leading to
\begin{equation}\label{5.5}
f_\pi^2 = 0.00929 GeV^2 = (96.4 MeV)^2 
\end{equation}
in rather close agreement with the standard value $93 MeV$, with no adjustable parameters.
The temperature correction is again proportional to $T^2$, which seems to be a rather general
result arising out of the first two terms in the contour integral representation, eq.(3.7),
a la Kislinger-Morley [16] of the Matsubara formalism [15], but we omit the result for brevity.  

\section{ Summary and Conclusion}

\setcounter{equation}{0}
\renewcommand{\theequation}{6.\arabic{equation}}

In retrospect, we have proposed a simple form of non- perturbative QCD, obtained by two 
differentiations of the o.g.e. propagator which, in conventional wisdom (almost since the 
inception of QCD), amounts to a Lorentz-invariant generalization of a $linear$ potential. 
[This is a rather qualitative statement, with not more than asymptotic validity, if one
expects a $quantitative$ behaviour like $1/k^4$ of the gluon propagator for finite $k^2$].  
The proposal has been motivated mainly by a desire for an alternative approach for understanding  non-perturbative QCD, since the predictions of some of the well-known `standard' methods like QCD-SR [2-4] and chiral perturbation theory [5,6] indicate considerable variations on the temperature dependence of QCD condensates. While the fuller implications of this proposal(which is characterized by a process
of differentiation w.r.t. a small mass parameter) are yet to emerge, we have offered some preliminary tests through the calculation of two basic QCD condensates,plus the pionic constant $f_\pi$, together with their temperature dependence. For these simple applications the non-perturbative proposal merely amounts to the replacement of the o.g.e. propagator $k^{-2}$ of perturbative QCD to $2\mu^2/k^4$, but more complex
amplitudes  would presumably need a more microscopic formulation in terms of the gluon fields themselves
at the level of the QCD Lagrangian. Such a formulation requires  each gluon field $A_\mu$ appearing in the Lagrangian to be replaced as follows:
\begin{equation}\label{6.1} 
A_\mu \rightarrow A_\mu [1 + \mu \partial_m ] 
\end{equation}
with a suitable limiting process $m \rightarrow 0$ on the (small) gluon mass $m$ after the
requisite differentiations on the o.g.e. propagators in the resulting equations of motion
and/or Feynman amplitudes have been performed. The second term is supposed to be an effective
substitute for an $exact$ QCD treatment,although at this stage it is not clear how much of the 
latter it accounts for, while avoiding `double counting' with the first. The simplest scenario is one 
in which the derivative term is substituted for the o.g.e. term in a lowest order Feynman diagram
for a specified process, somewhat akin to the Pagels-Stokar [22] `dynamical perturbation theory'.
Although the precise connection of this term to a $closed-form$ solution of the QCD 
Lagrangian is not yet in sight, the parameter $\mu$ (which sets the scale of the theory)  has,
by virtue of the structure of (6.1), the $potential$ for a deeper understanding of non-perturbative QCD,
on analogous lines to QCD-SR [2] or chiral perturbation theory [5]. One may also add, by way 
of precaution, that (at this stage) there is no tangible relation of this parameter with the QCD 
dimensional constant which must obey the constraints of renormalization group theory.  

For purposes of the present applications, the $gg$ interaction kernel is defined via fig.1, showing the relative contributions of `direct' (D), `exchange' (X) and `contact' (C) terms arising from lowest order QCD Lagrangian to give a gauge-invariant package, except for the replacement of the perturbative o.g.e. propagator $k^{-2}$ by $2\mu^2 k^{-4}$, corresponding to a non-perturbative (linear) form. The relation of the condensate to this kernel is shown in fig.2 via looping up of the gluon lines in two successive stages. This lowest order treatment in the gluon case is justified by the absence of a Schwinger mechanism [13] since, as explained in the introduction,  our  gluon does not couple to an axial current [12]. The $\partial_m^2$ method lends itself readily to the DR method [14], leading to the rather
simple formulae (3.6-7), where the identification of the scale parameter with the universal
Regge slope seems to provide a rather welcome accord with QCD-SR [3], although we are not inclined
at this stage no to attach much significance to this coincidence. In a similar way, the corresponding 
result on the $q{\bar q}$ condensate, has necessitated a prior derivation of the mass function
$m(p)$ via the solution of the SDE. This treatment has been admittedly approximate, but we have found a simple recursive procedure which converges rapidly. Here again the same universal constant $\mu$
has been employed, and the QCD-SR value [2] has been almost reproduced.

The more interesting result is the temperature dependence of the corresponding condensates which
shows  corrections of order $T^2$ for both the gluon and quark condensates, with a somewhat 
bigger coefficient for the former than for the latter. In this context, as already noted at the
end of Section 5, the order $T^2$ effect is a rather general feature of the Kislinger-Morley [16]
contour integral representation, eq.(3.7), of the Matsubara formalism [15] wherein, after
contour closing, the temperature dependence that arises from first two terms has proportionality 
to a $T^2$ behaviour in terms of the 3D integral over the momenta.  Now for the quark condensate 
the $O(T^2)$ correction, eqs.(4.8-10), agrees with the prediction of chiral perturbation theory 
($\chi PT$)[6], albeit with a slower variation, but the result for the gluon condensate is up against 
a whole spectrum of predictions! Even QCD-SR seems to predict both $T^2$ [23] and $T^4$ [3-4] 
variations, of which only one [23] agrees with our prediction. On the other hand, $\chi PT$   
seems to predict [24] an even stronger dependence on (`flat' rise with) $T$, viz., a 
$O(T^8/\pi^4)$ correction. In the absence of a `standard' criterion, it is difficult to decide 
on which of these alternatives corresponds to the truth, except that lattice calculations [25] 
predict an almost flat $<GG>$, with rise in temperature more in line with the $\chi PT$ 
prediction. This suggests that the gluon condensate, which is related to the anomaly of the energy 
momentum tensor $T_{\mu\nu}$, is $not$ the relevant order parameter for the description of chiral restoration with temperature [25] ! If the lattice criterion is taken seriously, it would seem 
to indicate a verdict for $\chi PT$ over QCD-SR [3-4, 23] predictions, at least one of which 
[23] agrees with our result. Corrections to QCD-SR have been suggested [26], arising from non-diagonal condensates [26] at finite temparatures, which may play a role in resolving the discrepancy [25]. 

As to the effect of this discussion on our result showing a $T^2$ dependence, there is a case
for inclusion of neglected effects, albeit within the contours of the Kislinger-Morley
representation, eq.(3.7). The first correction that is yet to be investigated is the effect of the quark-gluon interaction on the pure gluonic self-couplings whose dominance over the quark-gluon 
couplings stems from the relative strengths of the color Casimir factors. A related question 
concerns the role of the pion which is almost a separate identity in the $\chi PT$ scenario [5]. 
On the other hand, the logic of the present investigation does not permit the pion to play such 
a central role, (except as one of the many $q{\bar q}$ composites!), but this is $not$ to deny its 
strong dynamical link with the theory, one of whose many manifestations is the $correct$ 
prediction of $f_\pi^2$, vide eqs.(5.4-5), within the strict premises of the theory. Indeed the
chiral character of the pion is a natural consequence of the (vector-like) quark-gluon interaction 
in the QCD Lagrangian, a simple check being the prediction of its very small mass as a $dynamical$ consequence of such interaction. Although we have not carried out this check within the present 
formulation, analogous treatments with the same QCD premises, albeit with  harmonic confining 
interactions [10, 27], automatically predict a very small mass [10, 27] for this unique 
pseudoscalar composite, by virtue of the chiral character of the (vector-like) interaction. 
Finally, the status of the quantity $\mu$ at this stage is that of an effective scale parameter 
analogous to $f_\pi$ of $\chi PT$, and $not$ that of the QCD scale parameter a la renormalization 
group theory! A proper study of this quantity can only be done in the context of the structure of 
eq.(6.1) at the level of the gluon fields themselves. These and other related applications like 
the pion form factor (for further calibration) and three-gluon condensates are currently being studied, before extending the ideas to the full hadronic sector with its wider ramifications.

\section{Acknowledgement}
This work is supported in part by the Taiwan CosPA Project of Ministry of
Education (MoE 89-N-FA01-1-4-3) and in part by National Science Council
(NSC90-2112-M-002-028). One of us (ANM) is deeply indebted to Marcel Loewe for
a most informative communication on the temperature dependence of
gluon condensates.

\end{document}